\documentclass[epsf,usegraphicx]{mn2e}

\usepackage{xspace}

\title[The defining characteristics of IPs -- the case
of three candidate systems]
{The defining characteristics of Intermediate Polars -- the case
of three candidate systems}

\author[]
{Gavin Ramsay$^{1}$, Peter J. Wheatley$^{2}$, A. J. Norton$^{3}$, 
Pasi Hakala$^{4}$,  Darren Baskill$^{5,6}$\\
$^{1}$Armagh Observatory, College Hill, Armagh, BT61 9DG\\
$^{2}$Department of Physics, University of Warwick, Coventry, CV4 7AL\\
$^{3}$Department of Physics \& Astronomy, The Open University, Milton Keynes, 
MK7 6AA\\
$^{4}$Tuorla Observatory, University of Turku, V\"ais\"al\"antie 20, FIN-21500
Piikki\"o, Finland\\
$^{5}$Department of Physics and Astronomy, University of Leicester, Leicester,
LE1 7RH\\
$^{6}$Royal Observatory Greenwich, National Maritime Museum, Park Row, 
Greenwich, London, SE10 9NF
}

\begin{document}
\outer\def\gtae {$\buildrel {\lower3pt\hbox{$>$}} \over 
{\lower2pt\hbox{$\sim$}} $}
\outer\def\ltae {$\buildrel {\lower3pt\hbox{$<$}} \over 
{\lower2pt\hbox{$\sim$}} $}
\newcommand{\ergscm} {ergs s$^{-1}$ cm$^{-2}$}
\newcommand{\ergss} {ergs s$^{-1}$}
\newcommand{\ergsd} {ergs s$^{-1}$ $d^{2}_{100}$}
\newcommand{\pcmsq} {cm$^{-2}$}
\newcommand{\ros} {\sl ROSAT}
\newcommand{\chan} {\sl Chandra}
\newcommand{\xmm} {\sl XMM-Newton}
\def\rchi{{${\chi}_{\nu}^{2}$}}
\newcommand{\Msun} {$M_{\odot}$}
\newcommand{\Mwd} {$M_{wd}$}
\def\Mdot{\hbox{$\dot M$}}
\def\mdot{\hbox{$\dot m$}}
\newcommand{\teff}{\ensuremath{T_{\mathrm{eff}}}\xspace}

\maketitle

\begin{abstract}
Intermediate Polars (IPs) are a group of cataclysmic variables (CVs)
which are thought to contain white dwarfs which have a magnetic field
strength in the range $\sim$0.1--10MG. A significant fraction of the
X-ray sources detected in recent deep surveys has been postulated to
consist of IPs. Until now two of the defining characteristics of IPs
have been the presence of high (and complex) absorption in their X-ray
spectra and the presence of a stable modulation in the X-ray light
curve which is a signature of the spin period, or the beat period, of
the accreting white dwarf. Three CVs, V426 Oph, EI UMa and LS Peg,
have characteristics which are similar to IPs. However, there has been
only tentative evidence for a coherent period in their X-ray light
curve. We present the results of a search for coherent periods in
{\xmm} data of these sources using an auto-regressive analysis which
models the effects of red-noise. We confirm the detection of a
$\sim$760 sec period in the soft X-ray light curve of EI UMa reported
by Reimer et al and agree that this represents the spin period. We
also find evidence for peaks in the power spectrum of each source in
the range 100--200 sec which are just above the 3$\sigma$ confidence
level. We do not believe that they represent genuine coherent
modulations.  However, their X-ray spectra are very similar to those
of known IPs. We believe that all three CVs are {\sl bona fide}
IPs. We speculate that V426 Oph and LS Peg do not show evidence for a
spin period since they have closely aligned magnetic and spin axes. We
discuss the implications that this has for the defining
characteristics of IPs.

\end{abstract}


\section{Introduction}

Cataclysmic Variables (CVs) are interacting binaries in which a white
dwarf accretes material from a red dwarf secondary star through Roche
lobe over-flow.  The majority of CVs accrete via an accretion disc,
with the flow forming a strong shock at some height above the
photosphere of the white dwarf: this results in the emission of
X-rays. For those white dwarfs which have a magnetic field strength
above $\sim$10 MG (the {\sl polars} or {\sl AM Her stars}), the field
is strong enough to prevent the formation of an accretion disc and the
accretion flow gets directed towards its magnetic poles. The high
field also locks the rotation rate of the white dwarf to the binary
orbital period. For those systems with a magnetic field strength
between $\sim$0.1--10MG the accretion disc gets disrupted at some
distance from the white dwarf: these are called the {\sl Intermediate
Polars} (IPs).

\begin{table*}
\begin{center}
\begin{tabular}{lrrrrrrrrr}
\hline
Target & Date & MOS1 & MOS2 & pn & Duration & Count Rate\\
       &  & Mode & Mode & Mode & (sec) & (Ct/s)\\
\hline
EI UMa & 2002-11-02 & lw & sw & ff & 8246 & 6.6\\
LS Peg & 2005-06-08 & lw & lw & ff & 42535 & 0.70 \\
V426 Oph & 2006-03-04 & lw & sw & sw & 35964 & 5.5\\
\hline
\end{tabular}
\end{center}
\caption{The log for the observations of EI UMa, LS Peg and V426 Oph
made using {\xmm}. We show the mode the detector was in where `sw'
refers to `small window', `lw' large window, `ff' to `full frame'.
The duration of the observation refers to that using the EPIC pn
detector, while the Count Rate is determined over the 0.15--10keV
energy range.}
\label{log}
\end{table*}

Patterson (1994) reviewed the properties of then known IPs and noted
six which gave `significant clues' as to their IP nature. These
include a stable modulation in the optical and X-ray light curves
which is close to, or at, the spin period of the accreting white dwarf
($P_{spin}$); typically $P_{spin}<P_{orb}$; a side-band period which
results from a beat between $P_{spin}$ and $P_{orb}$; and a hard X-ray
spectrum which shows evidence of strong (and complex) absorption.  One
further X-ray property is the presence of strong Fe K$\alpha$ emission
(Norton, Watson \& King 1991).

At present only a relatively small number of IPs have been fully
identified. Patterson (1994) noted 18 {\sl bona fide} IPs, while Koji
Mukai's IP
homepage\footnote{http://asd.gsfc.nasa.gov/Koji.Mukai/iphome/iphome.html}
catalogues 29 confirmed IPs.  However, there is some evidence that IPs
are much more common than these numbers suggest. For instance, deep
X-ray observations of the Galactic centre using {\sl Chandra} led to
the discovery of a substantial population of hard X-ray sources which
Muno et al (2004) suggest are IPs.

It is therefore important to fully characterise the global properties
of the nearby CVs. Part of the difficulty lies in classifying CVs,
with individual systems showing characteristics of various sub-classes
of CV. Baskill, Wheatley \& Osborne (2005) performed a systematic
analysis of all the non-magnetic CVs observed using {\sl ASCA}. They
identified two systems, LS Peg, V426 Oph, which showed X-ray spectra
which were significantly more absorbed than the other CVs in their
sample. They also identified three systems, the previously noted and
EI UMa, which showed evidence for a periodicity in their X-ray light
curves (EI UMa also showed high absorption in its X-ray spectrum).

Baskill et al (2005) suggested that EI UMa, LS Peg and V426 Oph were
IPs. However, given the rather tentative nature of the periods which
they identified, further longer X-ray observations were required to
fully class these systems as {\sl bona fide} IPs. We have therefore
obtained {\xmm} data of LS Peg and V426 Oph -- observations of EI UMa
had already been performed in Nov 2002. We present an analysis of
these {\xmm} data and discuss the nature of these systems and how this
affects the defining characteristics of IPs. (A preliminary analysis
of the LS Peg data was presented by Baskill \& Wheatley 2006, while
Reimer et al 2008 report a period in the soft X-ray {\xmm} data of EI
UMa at $\sim$745 sec).

\section{Observations}

Observations were carried out using {\xmm} -- a log of the
observations is shown in Table \ref{log}. The EPIC X-ray detectors
(MOS, Turner et al 2001; pn, Str\"{u}der et al 2001) are relatively
wide-field instruments with medium spectral resolution. We do not
discuss the RGS detector data, with its higher X-ray spectral
resolution capability, nor do we discuss the observations made using
the Optical Monitor, since we are mainly concerned with X-ray timing
properties. Observations were also carried out of EI UMa in May 2002
but were affected by high background radiation.

The data were reduced using {\tt SAS} v7.1. Only X-ray events which
were graded as {\tt PATTERN}=0-4 and {\tt FLAG}=0 were used. Events
were extracted from a circular aperture centred on the source, with
background events being extracted from a source free area. The
background data were scaled to give the same area as the source
extraction area and subtracted from the source area. In the case of EI
UMa, the events in the EPIC pn detector (which was in full-frame mode)
were slightly piled-up during the time intervals which had the
greatest count rate. Since the MOS detectors were operating in small
window mode, the MOS data were not piled-up. For both LS Peg and V426
Oph we combined the light curves from the two EPIC MOS detectors and
compared the results obtained from the EPIC pn detector. We note that
in the case of LS Peg there was significantly enhanced background
12--20 ksec from the start of the observation.

\section{Light Curves}

\begin{figure*}
\begin{center}
\setlength{\unitlength}{1cm}
\begin{picture}(14,22)
\put(-0.5,15){\includegraphics{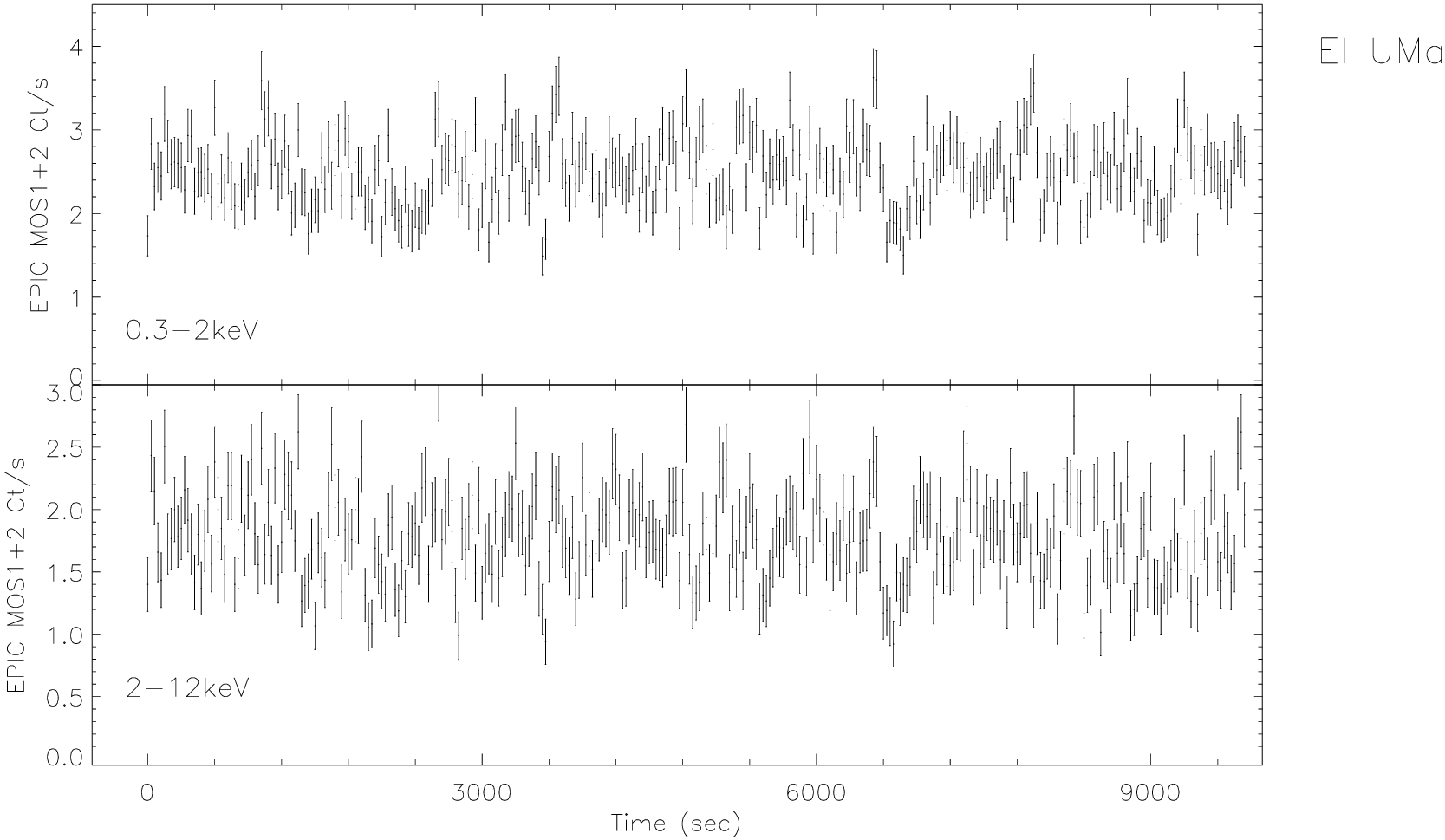}}
\put(-0.5,7.5){\includegraphics{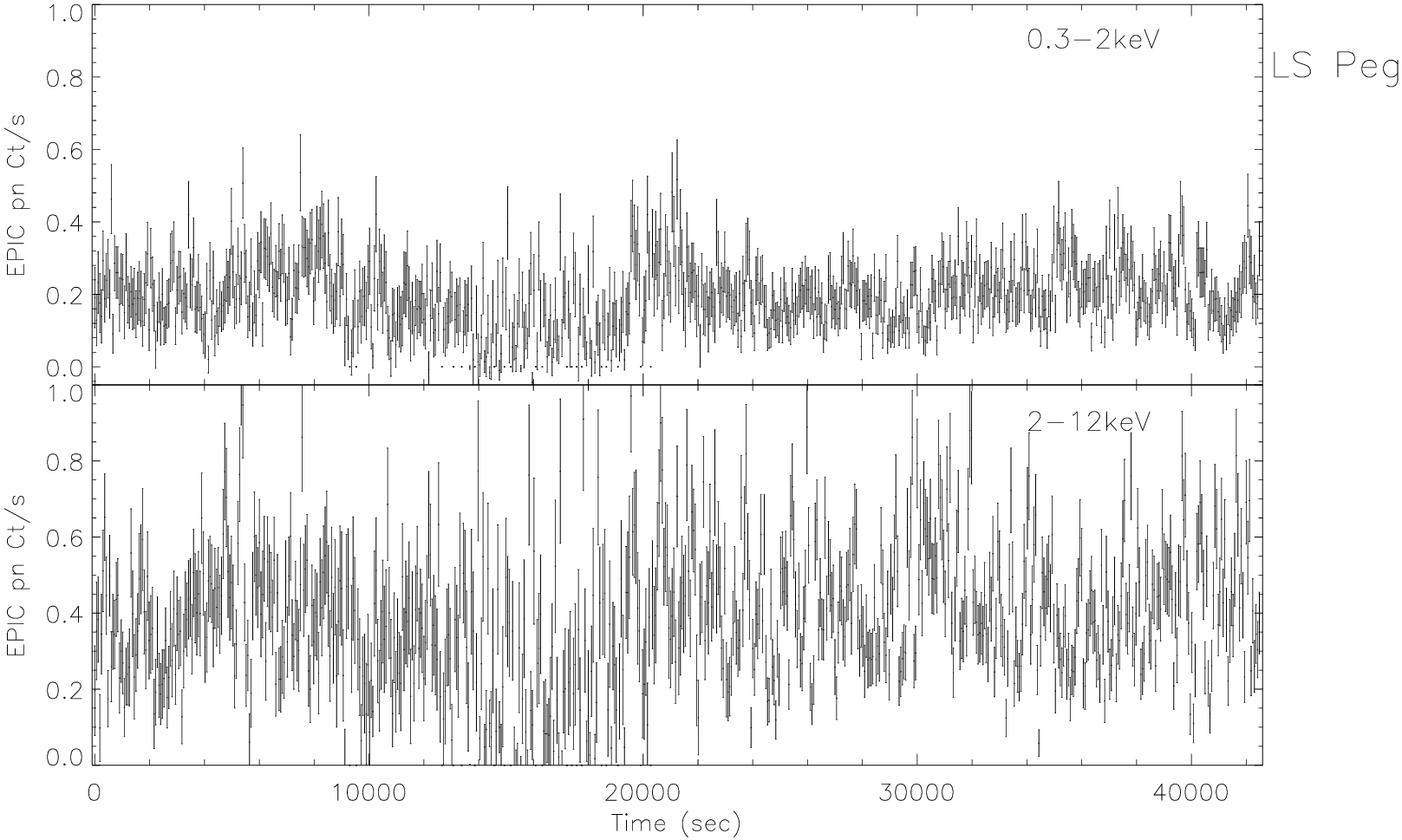}}
\put(-0.5,0.0){\includegraphics{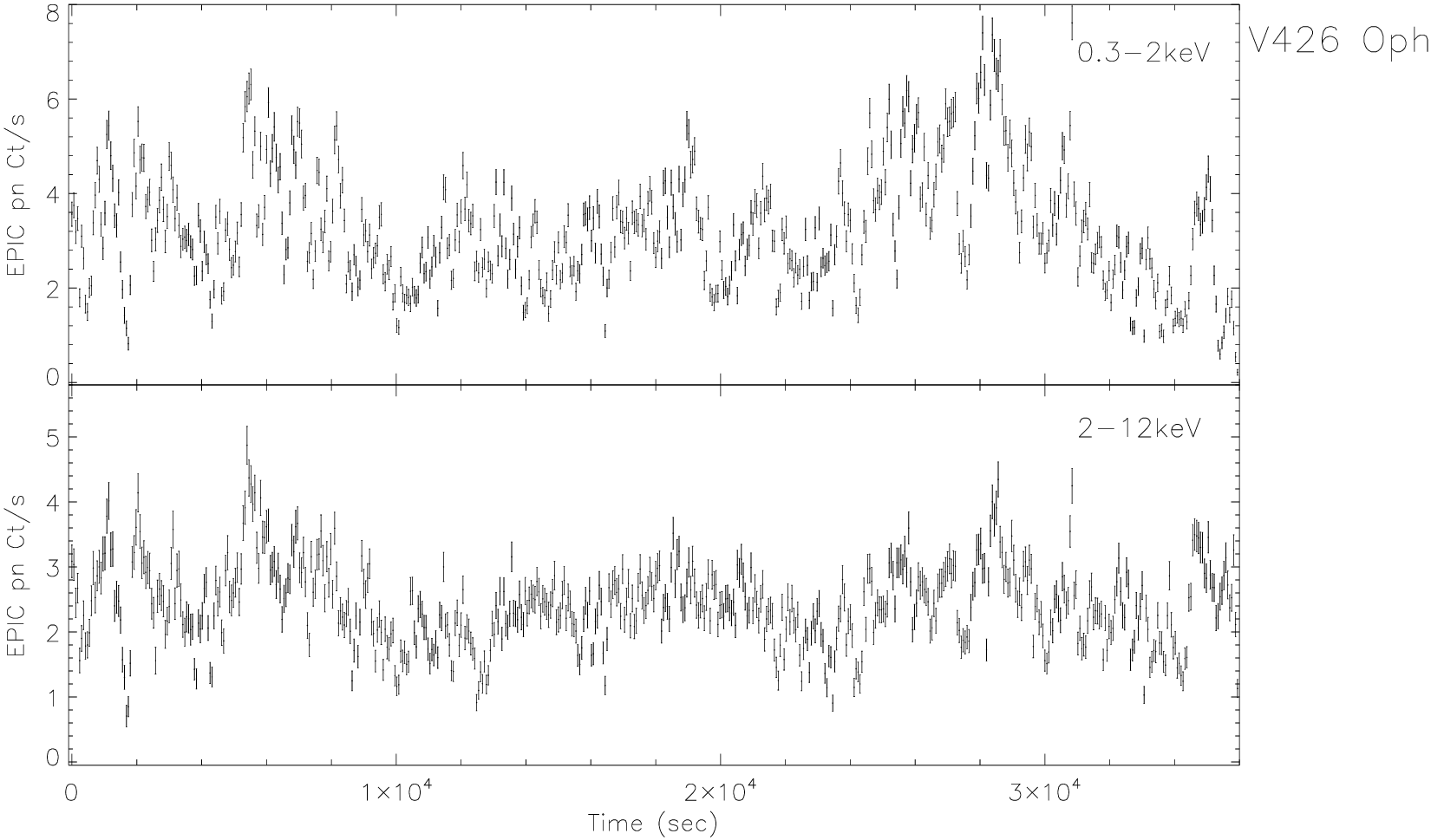}}
\end{picture}
\end{center}
\caption{From the top, the light curves of EI UMa (30 sec bins), LS
Peg and V426 Oph (60 sec bins). We show the soft (0.3--2 keV) and hard
X-ray (2--12 keV) light curves. In the observation of LS Peg the
particle/solar background was high between $\sim$12--20 ksec.}
\label{light}
\end{figure*}

We show the soft (0.3--2 keV) and hard (2--12keV) light curves of EI
UMa, LS Peg and V426 Oph in Figure \ref{light}. There is clear
evidence for variability in each of the light curves, although none
show variability on an obviously coherent timescale. V426 Oph shows
large variations in intensity on various timescales, particularly in
the soft X-ray band. The fact that the modulation in harder X-rays is
not as prominent suggests the soft X-ray variability is due to
variation in the absorption. In contrast, LS Peg shows less
variability in soft X-rays compared to hard X-rays.

\begin{figure*}
\begin{center}
\setlength{\unitlength}{1cm}
\begin{picture}(14,17)
\put(-3,5.5){\includegraphics{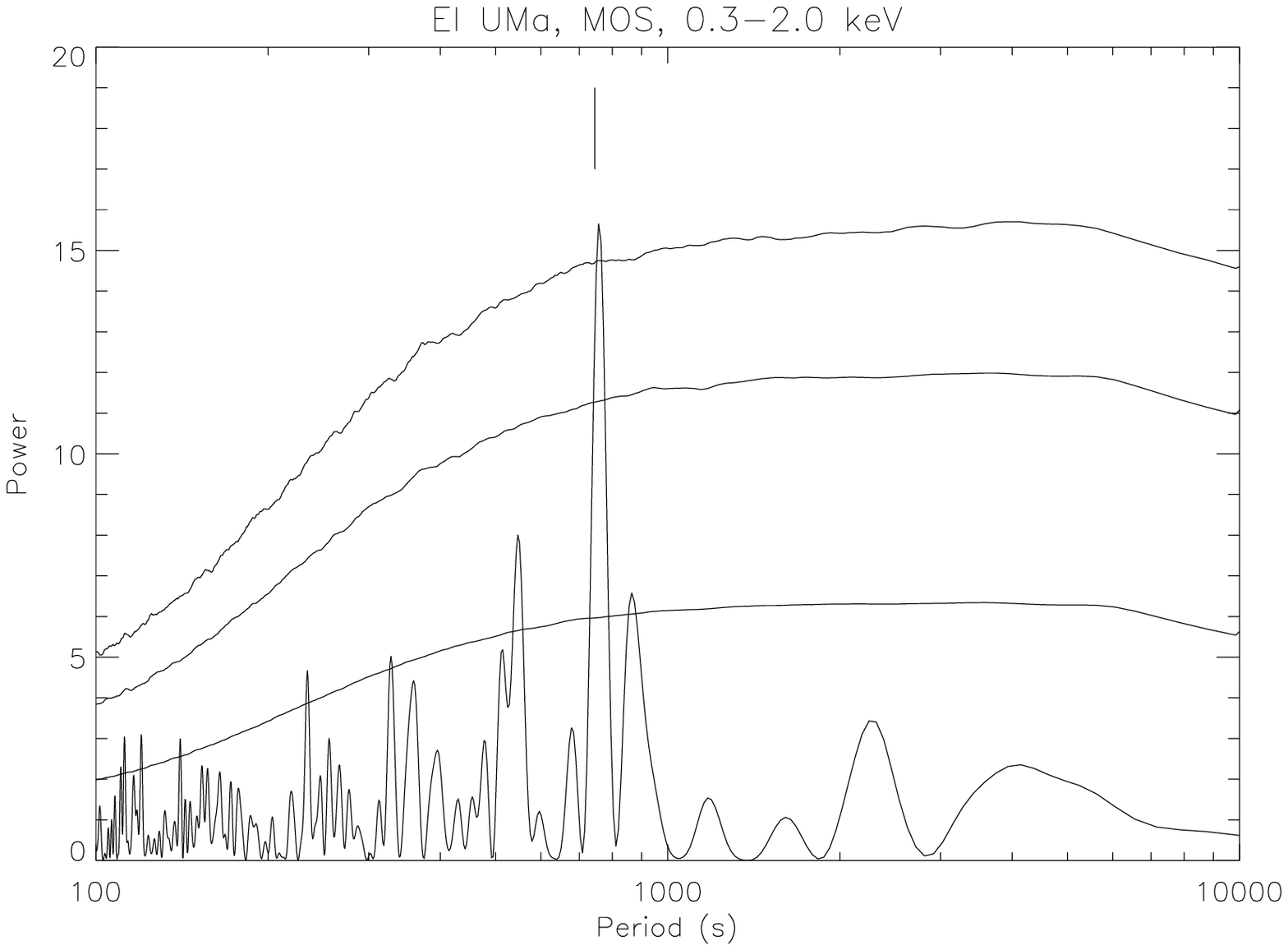}}
\put(5.5,5.5){\includegraphics{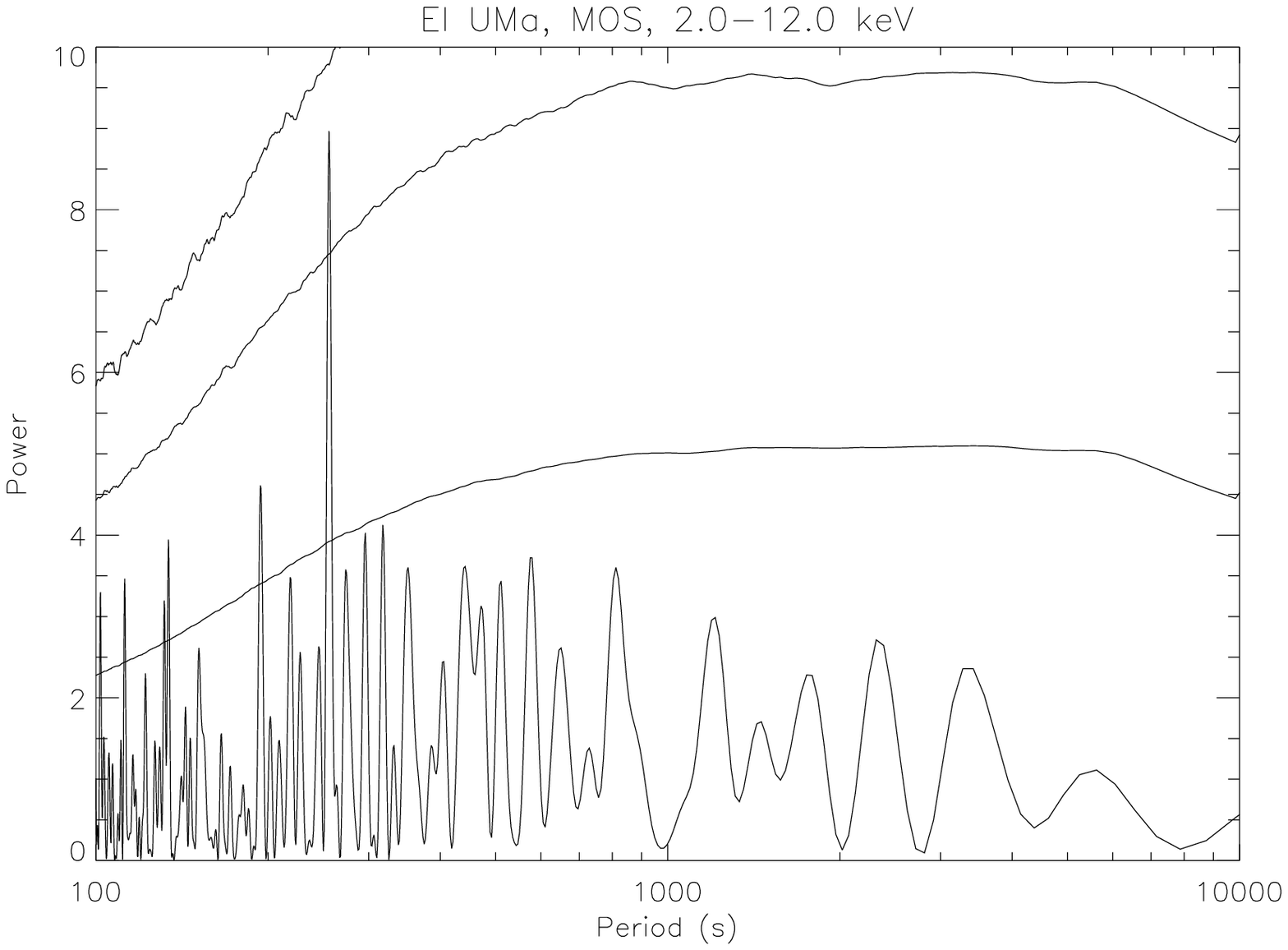}}
\put(-3,-0.5){\includegraphics{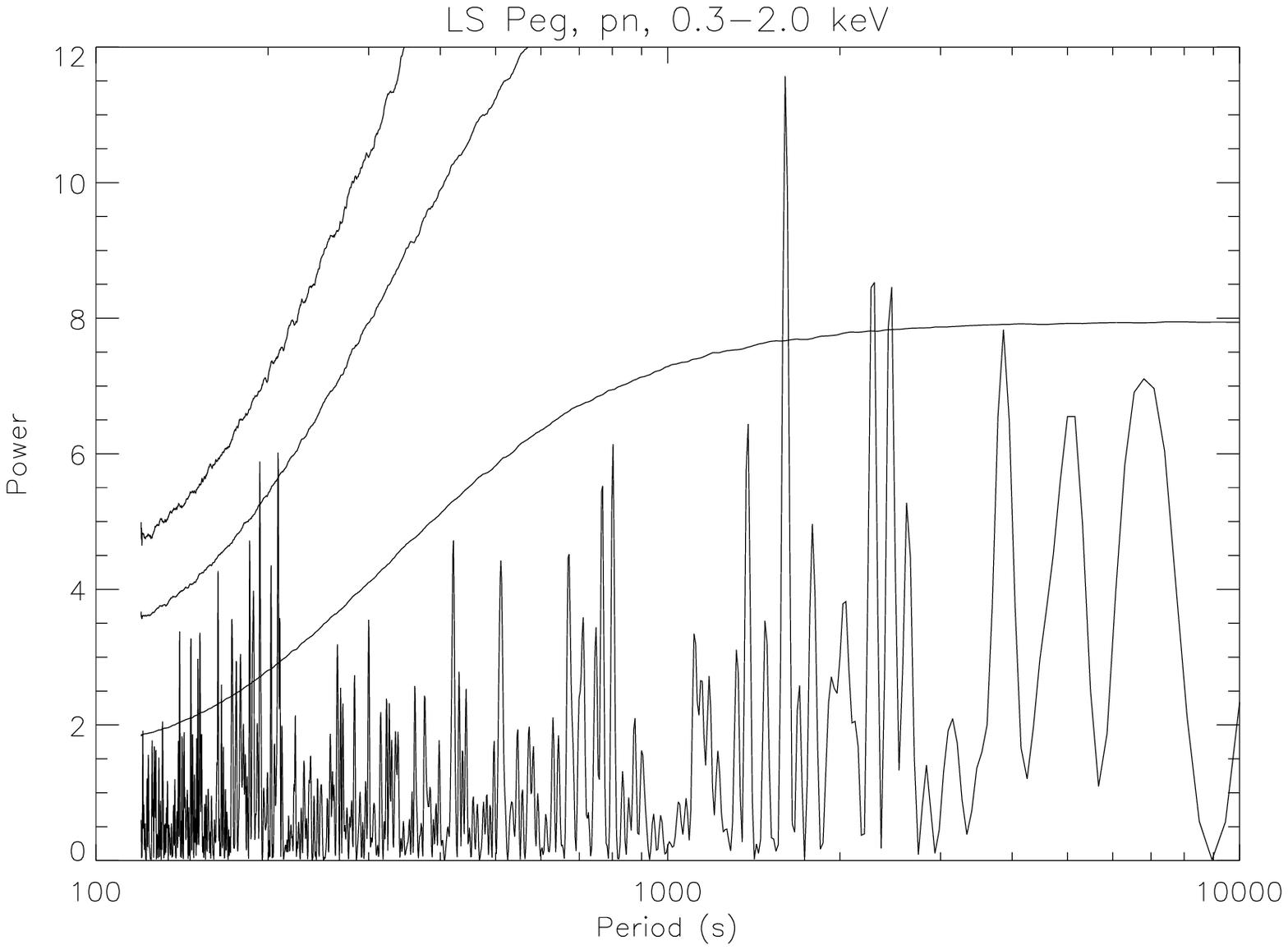}}
\put(5.5,-0.5){\includegraphics{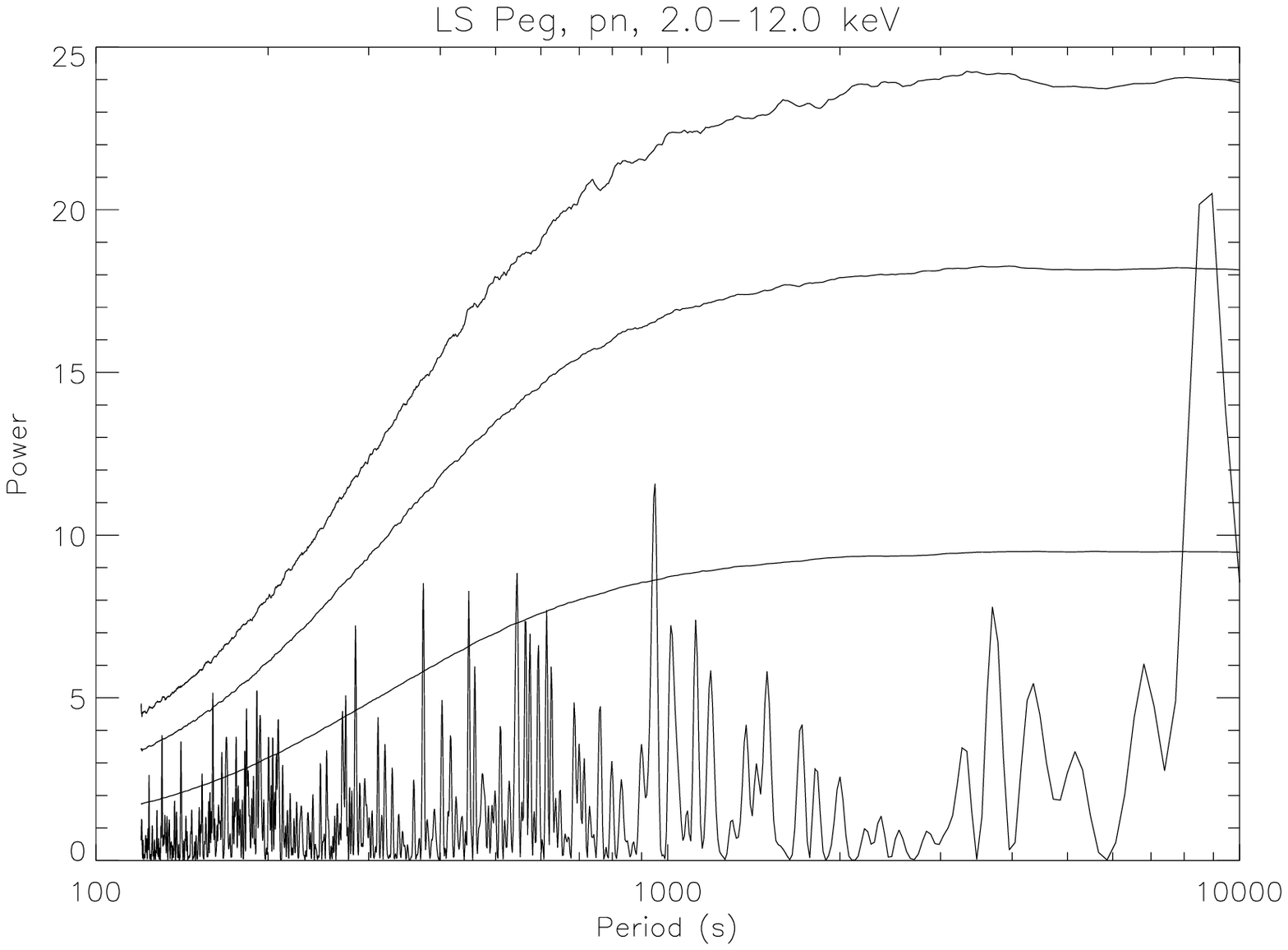}}
\put(-3,-6.5){\includegraphics{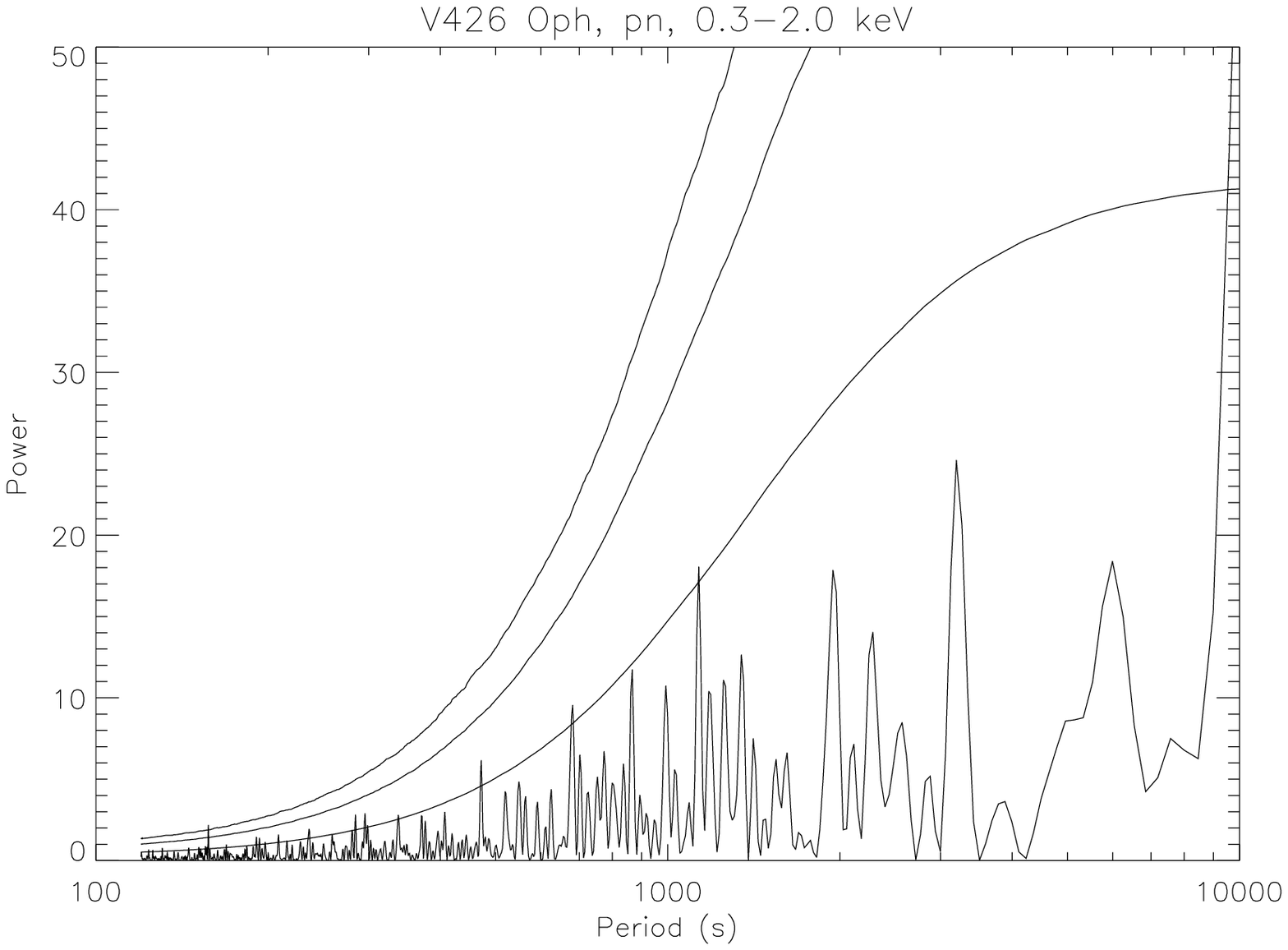}}
\put(5.5,-6.5){\includegraphics{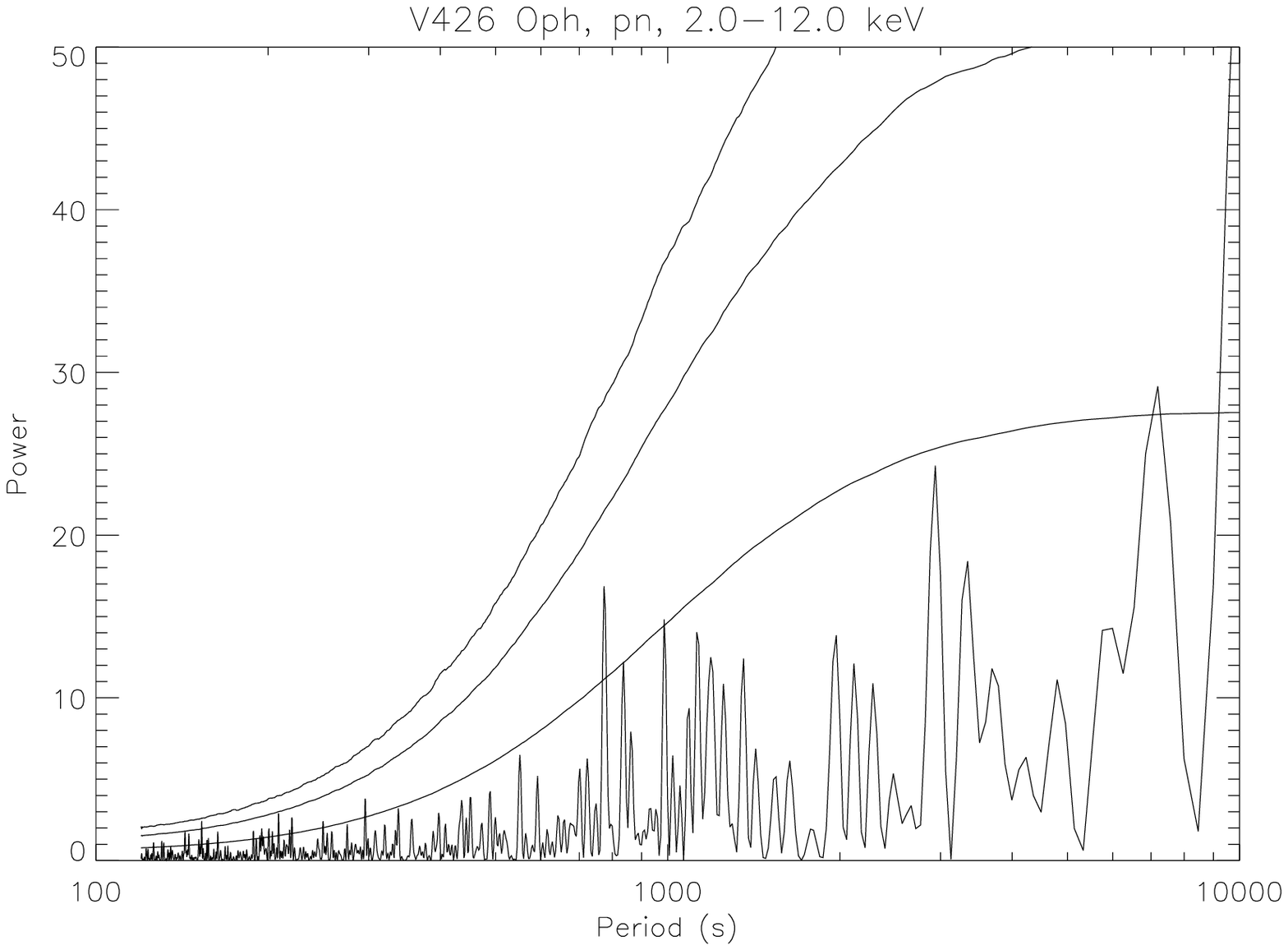}}
\end{picture}
\end{center}
\caption{From the top, the amplitude spectra of the soft (left hand
panels) and hard (right hand panels) X-ray light curves of EI UMa, LS
Peg and V426 Oph. We superimpose the confidence limits for power at
each period bin being significant at the 2$\sigma$, 3$\sigma$ and
4$\sigma$ confidence intervals. The `tick' mark in the top left panel
represent the period at which a modulated signal has previously been
reported (Table \ref{previous}).}
\label{power}
\end{figure*}

To search for evidence of periodic modulation in the light curves we
used the standard Lomb-Scargle power spectrum analysis. We show the
power spectra for the soft and hard X-ray light curves in Figure
\ref{power}. For light curves which are dominated by red noise (noise
dominated by low frequencies) non-standard techniques must be used to
determine the significance of peaks in the power spectra (which could
be due to red noise rather than strict coherent modulations). We used
the novel auto-regressive technique described in Hakala et al (2004).
Here we fitted a second order auto-regressive model to the data (ie
AR(2)). The fitted model is then used to simulate 10000 light curves,
which were in turn used to estimate the confidence limits. We
superpose the 95.2\% (2$\sigma$), 99.7\% (3$\sigma$) and 99.95\%
(4$\sigma$) confidence limits on the power spectra of each light curve
in Figure \ref{power}.

LS Peg shows a significant peak in its soft X-ray power spectrum which
corresponds to its orbital period (4.2 hrs). V426 Oph also shows a
peak in the soft X-ray power spectrum (6 hr) which is close to the
orbital period (6.85 hrs) and an additional peak (in both soft and
hard X-rays) at 3hrs. The orbital period of EI UMa (6.4 hrs) is
significantly longer than the duration of the observation. (Those
`long' periods are not shown in Figure \ref{power} since we are
searching for the shorter period modulations). In addition to these
periods, there are several peaks above the 3$\sigma$ confidence level,
but only one short period peak above the 4 $\sigma$ level, at
758$\pm$6 sec period in the 0.3--2 keV band light curve of EI
UMa. (This peak is also seen in the EPIC pn light curve which is
piled-up to some extent). There is no evidence for this period in the
2--12 keV light curve, although there is a period of $\sim$250 sec in
this higher energy power spectrum which has a significance of
$\sim3.7\sigma$. As noted by Reimer et al (2008) the X-ray period is
consistent with their detection of an optical period in the $U$ band
at 747 sec. Reimer et al also report the detection of other periods
between 760--770 sec which they take to be the beat period between the
orbital period and the spin period.

As the computed confidence limits correspond to the limits at
individual periods, there are bound to be some peaks above the
3$\sigma$ limit (1/370 probability) in the power spectra with hundreds
of independent frequencies -- this is indeed the case. In the case of
V426 Oph and LS Peg, none of the peaks with significance between
3--4$\sigma$ coincide with any previously reported periods, nor were
they seen in the power spectra derived using the other EPIC
instrument. We do not believe that these short period peaks represent
a genuine coherent modulation and therefore they do not represent the
spin period of the white dwarf.

\begin{table}
\begin{center}
\begin{tabular}{llll}
\hline
Target & Period &  & Ref \\
\hline
EI UMa   & 6.4 hrs & Orbital & 1\\
         & 12.4 min & X-rays & 2,3\\
         & 12.4--13.6 min & Optical & 3\\
LS Peg   & 4.2 hrs & Orbital & 4\\
         & 19 min & Optical & 5\\
         & 29.6 min & Circ Pol & 6\\
         & 30.9 min & X-rays & 2\\
V426 Oph & 6.85 hrs & Orbital & 7\\
         & :2.5, 4.5 or 12.5 hr & X-rays & 8\\
         & 28 min & X-rays & 9\\
         & 29.2 min & X-rays & 2\\
         & 2.1, 4.2 hrs & X-rays & 10\\
\hline
\end{tabular}
\end{center}
\caption{Previously reported periods in V426 Oph, LS Peg and EI UMa.
References: (1) Thorstensen (1986), (2) Baskill et al (2005), (3)
Reimer et al (2008), (4)
Taylor, Thorstensen \& Patterson (1999), (5) Szkody et al (2001), (6)
Rodriguez-Gil et al (2001), (7) Hessman (1988), (8) Rosen et al
(1994), (9) Szkody et al (1990), (10) Homer et al (2004).}
\label{previous}
\end{table}

\section{Spectra}

We give a brief overview of the characteristics of the {\xmm} spectra
of our three sources. Detailed spectral fits to the {\xmm} of EI UMa
has already been presented by Pandel et al (2005). They found that
good fits were only achieved using a complex absorption model (ie a
partial covering or an ionised absorption component was required in
addition to a single neutral absorption component). Further, the
column density of the partial covering absorption model was high,
$N_{H}\sim3\times10^{22}$ \pcmsq, and a covering fraction
$\sim$0.4--0.5. In contrast, the other non-magnetic CVs in the sample
of Pandel et al (2005) had a neutral absorption column of a few
$\times10^{20}$ \pcmsq.

We show in Table \ref{spec} the spectral parameters for all three
systems and we show the goodness of fit using a simple neutral
absorption component and also where we have added a partial covering
model.  We used the {\tt tbabs} neutral absorption model as used in
{\tt XSPEC} and also {\tt cemekl} which is a multi-temperature thermal
plasma model. We fitted the integrated X-ray spectrum and have not
attempted to determine if there is spectral variability over the
course of the orbital period -- which is likely to be the case. We
stress these are not definitive spectral fits for these objects --
rather we make the point that their X-ray spectra require complex
absorption components (as in the case of IPs) and their total
absorption column density is high compared to the non-magnetic
(non-eclipsing) dwarf novae.

We also show the equivalent width of the Fe K$\alpha$ line-complex at
6.4keV (due to fluorescence), $\sim$6.65 keV (the He-like line) and
$\sim$6.95 keV (the H-like line) in Table \ref{spec}. (The spectral
resolution of the EPIC detectors are not high enough to resolve the
He-like line into its three intrinsic components). We measured the
equivalent width by fitting an absorbed power law plus 3 Gaussian
components over the energy range 4--10 keV. While we stress that these
line results are intended to be informative rather than definitive,
the He-like line in all three systems show an equivalent width of
$\sim$200-500 eV. Both non-magnetic CVs (eg Rana et al 2006) and IPs
(eg Hellier \& Mukai 2004) typically show He-like lines with
equivalent widths of several 100 eV so their presence in our sources
is not conclusive either way.

\begin{table*}
\begin{center}
\begin{tabular}{lrrrrrrrrr}
\hline
Target & Single Absorber & $N_{H}$ & Multi Absorber & $N_{H}$ & $N_{H}$ & 
cvf & 6.4keV & 6.67keV & 6.95keV \\ 
       & (\rchi, dof) & \pcmsq & (\rchi, dof)& \pcmsq & \pcmsq & & (eV) & (eV)
& (eV) \\
\hline
EI UMa   & 3.60 (215) & 1.0$\times10^{20}$  & 1.80 (213) & 1.0$\times10^{20}$ &
4.5$\times10^{22}$ & 0.67  & 77 & 256 & \\
LS Peg   & 11.36 (444) & 2.2$\times10^{21}$ & 1.37 (441) & 1.7$\times10^{21}$ &
5.7$\times10^{21}$ & 0.91  & 118 & 478 & 86 \\
V426 Oph & 2.06 (530) & 1.8$\times10^{21}$ & 1.29 (528) & 1.0$\times10^{21}$ &
6.4$\times10^{21}$ & 0.65 & 34 & 302 & 153 \\
\hline
\end{tabular}
\end{center}
\caption{The best spectral fits to the {\xmm} EPIC spectra of our
three targets. We show the goodness of fit using a single neutral
absorption model and a neutral absorption component plus a partial
covering model (cvf). We used a multi-temperature thermal plasma model
and the emission model.}
\label{spec}
\end{table*}

\section{Is there evidence for the spin period in EI UMa, LS Peg and 
V426 Oph?}

Patterson (1994) showed that many IPs had $P_{spin}\sim0.1P_{orb}$.
This relationship predicts spin periods for our sources between
$\sim$1500--2500 sec. However, the current sample of IPs shows a wide
spread in $P_{spin}/P_{orb}$ -- see Figure 5 of Norton et al
(2008). Our {\xmm} light curves have allowed us to sample very low
values of $P_{spin}/P_{orb}$ (since our time-bins can be of short
duration). The limiting factor for higher values of $P_{spin}/P_{orb}$
is set by the time duration of the light curves.

Our red-noise analysis shows periods between 100--200 sec in the X-ray
light curves of each of our sources at a significance level just
greater than the 3$\sigma$ level. None of these periods have been
detected in previous X-ray data. On the other hand, apart from the
signature of the orbital periods and the 760 sec modulation in the
soft X-ray band of EI UMa reported by Baskill et al (2005) and Reimer
et al (2008), none of the previously reported periods (Table
\ref{previous}) were detected. For V426 Oph and LS Peg which were
observed for longer than 35 ksec we find no evidence for a spin period
between a few 100 sec and $\sim$10 ksec. In the case of EI UMa our
light curve is 8.2 ksec in duration so therefore cannot rule out the
presence of a period greater than this.

\section{The nature of EI UMa, LS Peg and V426 Oph}

Hellier et al (1990) discussed the possibility of V426 Oph being an IP
and concluded it was not since there was no prominent X-ray
modulation. In the case of LS Peg, the fact that there is some
evidence that its optical flux is circularly polarised (albeit only at
a significance level of 3$\sigma$) on a period of 29 min hints at a
magnetic nature (Rodriguez-Gil et al 2001). Indeed, a number of other
IPs show a modulation in the optical circular polarisation at the spin
period (see the Katajainen et al 2007 for a recent reference list).
It is clear that there is a great deal of uncertainty as to the
sub-type of V426 Oph and LS Peg. For instance, in the catalogue of
Ritter \& Kolb (2003), LS Peg is classed as an nova-like, while VY
Scl, SW Sex and IP! In the case of EI UMa both Pandel et al (2005) and
Reimer et al (2008) suggest it is an IP.

We believe that the fact that all three systems show high levels of
complex absorption (ie an equivalent column density $>5\times10^{21}$
\pcmsq, Table \ref{spec}) point to these systems being {\sl bona fide}
IPs. We now address why they do not show evidence for a spin period in
X-rays.

\section{The defining characteristics of IPs}

A strong X-ray modulation has always been taken as one of the main
defining characteristics of an IP. A modulation at the spin period
arises due to phase-varying photoelectric absorption in the accretion
column and/or self occultation by the body of the white dwarf. A
similar modulation at the beat period arises due to the accretion
stream flipping between the poles.  However, these modulations will
only arise if the magnetic axis of the white dwarf is mis-aligned with
the white dwarf spin axis (eg King \& Shaviv 1984). Therefore, for all
confirmed IPs, it is most likely that the magnetic and spin axes are
offset from each other, typically by $\sim 10$'s of degrees. Only in
one IP (XY Ari) is there an observational measurement of this offset:
Hellier (1997) estimates that in XY Ari the magnetic dipole axis is
offset from the spin axis by 8--27$^{\circ}$ and that the main
accretion region is offset from the magnetic axis by a further
19$^{\circ}$.

If the magnetic and spin axes of a white dwarf in an accreting binary
system are closely aligned, then a disc-fed system will give rise to
circular accretion curtains above each magnetic pole (extending to all
azimuths).  There will essentially be no phase varying photoelectric
absorption and no variation in self occultation, and consequently the
X-ray signal will show no variation with spin phase. Similarly, in a
stream-fed system with co-aligned magnetic and spin axes, equal
amounts of material are likely to feed onto each pole at all times,
and there will be no variation in X-ray flux with beat phase. Axes
that are aligned to within a few degrees will likely give rise to an
X-ray modulation of less than a few percent depth, which would be
undetectable in most cases due to intrinsic X-ray flickering. In
contrast, simulations made using the polarisation models of Potter,
Hakala \& Cropper (1998) show that a mis-alignment between the spin
and magnetic axes of even 5$^{\circ}$ (and assuming accretion onto a
small spot at the magnetic poles) gives rise to a modulation in the
circular polarisation flux as seen in LS Peg.

We therefore suggest that there should exist a population of IPs with
closely aligned spin and magnetic axes which do not show any
detectable spin or beat modulation in their X-ray flux. Should such
systems exhibit any other defining variation in their X-ray flux?

Up to half of all confirmed IPs also show an X-ray modulation at the
system orbital period (Parker, Norton \& Mukai 2005). This is presumed
to be due to absorption in material at the edge of the accretion disc,
thrown up by the impact of the accretion stream. Since $\sim$50\% of
all known systems show this effect, and assuming their orbital planes
to be randomly orientated with respect to our line of sight, this
indicates that the absorbing material must extend far enough out of
the orbital plane to be seen at all inclination angles above
$60^{\circ}$. The question arises therefore, should not half of the
IPs with aligned spin and magnetic axes show such an orbital
modulation and be detected that way?

Norton \& Mukai (2007) showed that the IP XY Ari exhibits a broad
X-ray modulation at its orbital period which comes and goes on a
timescale of $\sim$months. They interpreted this as evidence for a
precessing, tilted or warped accretion disc in that system. They
further suggested that such discs may exist in all IPs and that the
cause of the warp may be related to the spinning magnetic field
anchored to the white dwarf at its centre. A tilted magnetic dipole
will stir up the inner edge of the accretion disc and lift material
away from the orbital plane, thus initiating the precessing warp or
tilt. The lifting of material far enough out of the orbital plane to
cause an X-ray orbital modulation in half the known IPs is therefore
due to this warp or tilt in the disc, which in turn is due to the
tilted dipole at its centre, caused by mis-aligned spin and magnetic
axes.

For IPs with closely aligned spin and magnetic axes, the situation is
less clear. However, we suggest that for these systems, material is
azimuthally concentrated at the point where the accretion stream from
the secondary star meets the accretion disc, thereby giving rise to a
modulation at the orbital period. Closer to the white dwarf, the
accretion flow is more azimuthally symmetric and hence there is no
modulation at the spin period and no precessing, tilted or warped
accretion disc.

\section{Conclusions}

We have analysed the X-ray light curves of EI UMa, LS Peg and V426 Oph
using a procedure which models the effect of red-noise in an
appropriate manner. We have found evidence for peaks in the power
spectra between $\sim$100-200 sec which are marginally above the
3$\sigma$ confidence level but only one peak above 4$\sigma$ (760 sec
in the soft X-ray light curve of EI UMa). Apart from the latter peak,
these periods have not been detected in previous observations of these
sources, nor have we detected significant peaks in the power spectra
at previously noted periods. We find no evidence for a modulation on
the spin period in either LS Peg or V426 Oph, but confirm the result
of a 760 sec period in EI UMa as reported by Baskill et al (2005) and
Reimer et al (2008) which they believed is the signature of the spin
period.

However, since all three systems show complex and high absorption we
believe that they are {\sl bona fide} IPs. We speculate that we do not
detect an X-ray modulation in LS Peg and V426 Oph at their spin period
since the spin and magnetic axes of the accreting wide dwarf are
closely aligned. V426 Oph and LS Peg show evidence for an orbital
modulation in soft X-rays since material is lifted out of the orbital
plane and crosses our line of sight to the accretion regions on the
white dwarf.

Therefore, we speculate that there is a subset of IPs which are not
expected to show a short period coherent modulation -- previously
taken to be a necessary characteristic of IPs. This will have
implications for identifying objects as IPs and hence their space
density. For instance, Barlow et al (2006) present a compilation of
CVs detected using {\sl Integral}/IBIS survey data. Out of the 9 CVs
discovered as a result of {\sl Integral} data, 4 have no detected
period and are therefore regarded as `unclassified' CVs. The fact that
they are detected at hard X-rays (20--100keV) is consistent with them
being IPs. We suggest that they are {\sl bona fide} IPs which have
closely aligned spin and magnetic axes and therefore do not show a
coherent X-ray signal on their spin period.

\section{Acknowledgements}

Armagh Observatory is grant aided by the N. Ireland Dept. of Culture, Arts 
and Leisure. We thank Stephen Potter for advice on simulating polarisation
light curves of IPs.

\end{document}